\documentclass[]{aastex6}

\fullcollaborationName{The Friends of AASTeX Collaboration}

\begin{document}


\title{Evolution of High-Energy Particle Distribution in Mature Shell-Type Supernova Remnants}


\author{Houdun Zeng\altaffilmark{1,3}, Yuliang
Xin\altaffilmark{1,4}, Siming Liu\altaffilmark{1}, J.  R. Jokipii\altaffilmark{5}, Li Zhang\altaffilmark{2,3}, and Shuinai Zhang\altaffilmark{1}}


\altaffiltext{1}{Key Laboratory of Dark Matter and Space Astronomy,
Purple Mountain Observatory, Chinese Academy of Sciences,
Nanjing 210008, China, zhd@pmo.ac.cn; liusm@pmo.ac.cn}
\altaffiltext{2}{Department of Astronomy, Yunnan University, Kunming, 650091, China}
\altaffiltext{3}{Key Laboratory of Astroparticle Physics of Yunnan Province,
Kunming, 650091, China}
\altaffiltext{4}{University of Chinese Academy of Sciences, Yuquan Road 19,
Beijing,100049, China}
\altaffiltext{5}{University of Arizona, Tucson,
Arizona, 85721, USA}

\begin{abstract}

Multi-wavelength observations of mature supernova remnants (SNRs), especially with recent advances
in $\gamma$-ray astronomy, make it possible to constrain energy distribution of energetic particles
within these remnants. In consideration of the SNR origin of Galactic cosmic rays and physics related
to particle acceleration and radiative processes, we use a simple one-zone model to fit the nonthermal
emission spectra
of three shell-type SNRs located within 2 degrees on the sky: RX J1713.7$-$3946, CTB 37B, and CTB 37A.
Although radio images of these three sources all show a shell (or half-shell) structure, their radio,
X-ray, and $\gamma$-ray spectra are quite different, offering an ideal case
to explore evolution of energetic particle distribution in SNRs. Our spectral fitting shows that 1)
the particle distribution becomes harder with aging of these SNRs, implying a continuous acceleration
process, and the particle distributions of CTB 37A and CTB 37B in the GeV range are harder than the
hardest distribution that can be produced at a shock via the linear diffusive shock particle acceleration
process, so spatial transport may play a role; 2) the energy loss timescale of electrons at the
high-energy cutoff
due to synchrotron radiation appears to be always a bit (within a factor of a few) shorter than the
age of the corresponding remnant, which also requires continuous particle acceleration; 3)
double power-law distributions are needed to fit the spectra of CTB 37B and CTB 37A, which may
be attributed to shock interaction with molecular clouds.
\end{abstract}

\keywords{cosmic rays - gamma rays: ISM - ISM: supernova remnants - radiation mechanisms: non-thermal}



\section{Introduction} \label{sec:intro}

Supernova remnants (SNRs) have been considered as the dominant cosmic ray sources in
the Galaxy since 1930s. Direct observational evidence for this paradigm comes from
detection of radio and X-ray emission produced via the synchrotron process by GeV and
TeV electrons, respectively, and $\gamma$-ray emission produced via three processes:
inverse Compton (IC) scattering of low energy photons in the background
and bremsstrahlung by relativistic electrons (leptonic processes), and decay of neutral
pions produced in the inelastic hadronic interaction of high-energy ions with ambient nuclei
(hadronic process)\citep {Vink12}. Determining the nature of the $\gamma$-ray emission
plays an essential role in evaluating contributions of SNRs to the flux of Galactic cosmic
rays observed near the Earth \citep{Acero2016}. Thanks to their excellent sensitivities,
Cherenkov telescopes on the ground and the Large Area Telescope (LAT) on board the Fermi
satellite have detected a few tens of supernova remnants with good spectral measurement in
the $\gamma$-ray range during the past decade \citep{Ferrand2012,Carrigan2013,Acero2016},
revealing a variety of $\gamma$-ray spectra \citep{Yuan2012}.
Moreover analyses of \textit{Fermi}-LAT and AGILE observations of SNRs W44, IC443 \citep{Tavani2010, Giuliani11,Ackermann2013},
and W51C \citep{Jogler2016} have revealed tentative evidence for a low-energy spectral turnover associated with the $\pi^0$ decay process.


The SNR complex CTB 37, containing CTB 37A (G348.5$+$0.1), CTB 37B (G348.7$+$0.3) (and G348.5$-$0.0,
which is not detected in the $\gamma$-ray range and will not be studied here),
is an interesting region for comparative study of SNRs.
The bright TeV SNR RX J1713.7$-$3946 (HESS J1713$-$397, also known as G347.3$-$0.5) locates within
one degree of this complex \citep{Aharonian2006, Aharonian2008a}.
CTB 37A and 37B were discovered at radio wavelengths with very similar surface brightness ($\sim $ 
$8.5 \times 10^{-20}~\rm W m^{-2} Hz^{-1} sr^{-1}$ at 20 cm), sizes ($\sim 10^{'}$), and
spectral indices ($\alpha \sim 0.3-0.5$) \citep{Clark1975, Kassim1991}.
At TeV energies, HESS J1713$-$381 and J1714$-$385 have been identified as the counterpart of
CTB 37B and 37A, respectively \citep{Aharonian2008a}.
All these three remnants have a central compact object and show a prominent shell (or half-shell)
structure caused by shocks driven by core collapse supernovae. However, while the diffuse X-ray
emission from RX J1713.7$-$3946 is dominated by synchrotron emission of TeV electrons \citep{Katsuda15},
the diffuse X-ray emission from CTB 37A and CTB 37B are predominantly thermal. 
In the GeV range, RX J1713.7$-$3946 has a very hard spectrum, typical for young shell-type SNRs
with prominent synchrotron X-ray emission \citep{Acero2015}. CTB 37A has a very soft spectrum \citep{Brandt2013},
typical for SNRs interacting with molecular clouds. The GeV spectrum of CTB 37B is a bit
peculiar with a spectral peak near $\sim 10$ GeV \citep{Xin2016}. The above characteristics
make these three SNRs to be ideal candidates for a comparative study of nonthermal emission from
mature SNRs driven primarily by strong collisionless shocks to explore the evolution of
energetic particle distribution in SNRs. Considering the SNR origin of Galactic cosmic rays,
we propose a unified model with as few parameters as possible for the spectral fitting with
the Markov Chain Monte Carlo (MCMC) algorithm \citep{Lewis2002}. The derived model parameters
can be used to study the underlying physical processes of strong astrophysical shocks.


Our model is described in Section 2. In Section 3, we show 
results of the spectral fitting and compare them with previous studies.
The conclusion and discussion are presented in Section 4.

\section{Model Description} \label{sec:model}

The physical processes of charged relativistic particle acceleration by inductive electric fields
only depend on the particle rigidity $R=p/q$ explicitly, where $p$ and $q$ are the momentum and charge of
the particle, respectively \citep{Zhou2016}. However the energy loss of high-energy electrons and ions
via interactions with the background plasma are quite different. The radiative energy loss of electrons
can make its energy distribution cutoff at an energy much lower than that for ions. We will assume that
the distribution function of all particles has the following form
\begin{equation}
N(R_i)=N_{0,i} R_i^{-\alpha}\left(1+\frac{R_i}{R_{\rm br}}\right)^{-1}\textrm{exp}\left(\frac{-R_{i}}{R_{i,{\rm cut}}}\right) \Theta(R_i-R_{\rm min})
\end{equation}
where $R_i = p_i/q_i$ and '$i$' represents different particle species, $\Theta$ is the Heaviside function,
which is zero for negative argument and 1 for positive argument. For $R_{\rm min} < R_{\rm br} \ll R_{i, {\rm cut}}$,
we have a double power-law distribution with an exponential high-rigidity cutoff.
For $R_{\rm min} < R_{i, {\rm cut}} < R_{\rm br}$, the distribution function is approximately a single power-law
with an exponential cutoff. Different particle species therefore have the same spectral break $R_{\rm br}$ and index $\alpha$,
but their high-rigidity cutoffs $R_{i, \rm cut}$ may be different. Motivated by the mechanism for
formation of spectral breaks proposed by \citet{Malkov11} and to simplify the model, we assume that
for the broken power-law distribution, the spectral index increases by one toward higher rigidities. 
We set $R_{\rm min}=1~{\rm MeV}/ce$, where $c$ and $e$ are the speed of light and the elementary charge unit,
respectively, so that electrons are still relativistic and can produce synchrotron emission efficiently.
Considering the SNR origin of Galactic cosmic rays, we will assume that $K_{ep}=N_{0, e}/N_{0, p}=0.01$
\citep{Yuan2012} and the radiative effect of ions heavier than protons are included in $N_{0, p}$.
The total energy content of protons above 1 GV$=1$ GeV$/ec$ then determines the normalization of the particle distributions.

With the relativistic particle distributions given above, the nonthermal radio to X-ray data can
be fitted with the electron synchrotron processes \citep{Ghisellini1988,Strong2000} with the mean magnetic
field $B$ as a free parameter. For $\gamma$-ray emission, IC scattering of background
soft photons by electrons \citep{Jones1968}, electron bremsstrahlung \citep{Strong2000} and decay of
neutral pions produced via proton-proton inelastic collisions are considered \citep{kamae2006}.
For the electron bremsstrahlung and hadronic processes, we assume that the background plasma has the same density.
Then above $\sim 100$ MeV, the $\gamma$-ray flux produced via $\pi^0$ decay is always higher
than that produced via the bremsstrahlung.
For the IC scattering of electrons, besides the cosmic microwave background radiation (CMB),
we include an infrared field with $T=30 $ K and an energy density of $1$ eV cm$^{-3}$ \citep{Porter2006}.
The distance, source size, age, and the gas density in the emission region can be
obtained from observations. The value of distance to an SNR is an extremely
important quantity, which determines the size and energy contents in energetic
particles and the magnetic field. Unfortunately, it is usually quite difficult to
determine the distance precisely \citep{Acero2016}. These distances are often estimated
as the kinematic distance of associated molecular clouds detected via CO lines
(e.g. \citealp{Castelletti2013}), OH maser or HI absorption features (e.g. \citealp{Tian2011,Tian2014}).
The extinction column density of optical and/or X-ray emission and/or the dispersion
measure of the associated radio pulsar can also be use to constrain the
distance (e.g. \citealp{Arzoumanian2011}).
For shocks interacting with molecular clouds, given the complexity of this interaction process
\citep{Jones93,Inoue2012,Fukui2012,Sano2010,Sano2015}, we will adjust the effective density
for the emission process so that the total energy content in relativistic
protons is on the order of $10^{50}$ erg, as required by the scenario of SNR origin of Galactic cosmic rays.
Note that, since we are mainly interested in the overall emission spectrum, the effective density partly
takes into account inhomogeneity of the ISM and uncertainties pertaining to interaction of cosmic rays
with molecular clouds \citep{Reach2005,Fukui2012,Sano2010,Sano2015}.

Therefore, there are only 5 free parameters: $\alpha$, $R_{\rm e, cut}$, $R_{\rm p, cut}$, $N_{0\rm, p}$, and
the magnetic field for synchrotron emission $B$. Whenever necessary, a spectral break $R_{\rm br}$ is also introduced.
Considering the radiative cooling effect of high-rigidity electrons, the synchrotron cooling
timescale of electrons near $R_{e,cut}$,
\begin{equation}
\tau_{\rm syn} = 1.25 \times 10^3 \left(\frac{R}{1\rm TV}\right)^{-1}\left(\frac{B}{100 \mu {\rm G}}\right)^{-2} \,\, \rm{year}\,,
\end{equation}
should be longer than the diffusive shock acceleration timescale with Bohm diffusion \citep{Lagage1983,Drury1991},
\begin{equation}
\tau_{\rm acc}=20 \left(\frac{c^2 R}{3B U^2}\right) = 20 \left(\frac{R}{1 \rm TV}\right) \left(\frac{B}{100\mu {\rm G}}\right)^{-1}\left(\frac{U}{10^{3} km/s}\right)^{-2} \,\, \rm{year}\,,
\end{equation}
where 1TV = 1TeV$/ec$.
Then we have
\begin{equation}
R_{\rm e, cut} < 8 \left(\frac{U}{10^3 \rm km/s}\right)\left(\frac{B}{100\mu G}\right)^{-1/2} \,\, {\rm TV}\,,
\end{equation}
where $U$ is the shock speed. Note that for perpendicular shocks, the acceleration rate can be much higher, which will lead to a shorter acceleration timescale and a higher upper limit for the rigidity \citep{Jokipii1987}.
We will use the MCMC technique to constrain the model parameters. To find the best fit to observational
data in a multi-dimensional model parameter space, the MCMC method is widely used for its high efficiency.
In this approach, a Markov chain is built with the Metropolis-Hastings sampling algorithm that determines
the jump probability from one point to another in the parameter space. For each parameter set ${\bf P}$,
one obtains the likelihood function ${L(\bf P)}\propto {\rm exp}(-\chi^2(\bf P)/2)$, where $\chi^2$ is obtained by
comparing model predictions with observations. A new set of parameter ${\bf P^\prime}$ is adopted to replace
the existing one ${\bf P}$ with a probability of min\{$1, L({\bf P^\prime})/L({\bf P})$\}. This
sampling method ensures that the probability density distributions of model parameters are asymptotically
approached with the number density of sampling points. The MCMC method has been reviewed by
\cite{Liu2012} and described in detail by \cite{Neal1993,Gamerman1997,Lewis2002,Mackay2003}.
Since this paper focuses on the overall emission spectra of SNRs, the magnetic field derived via
the spectral fit should be interpreted as the spatially averaged value, similar to the effective
density we adopted. Detailed studies of spatial structure of individual SNRs (e.g., \citealp{Sano2015})
are highly complementary to our results.
For sources modeled with different effective densities, the Akaike information criterion, ${\rm AIC} = -2\ln L+2k$, where $k$ is the number of model parameters \citep{Liddle2007}, can be used to determine statistically which model is preferred by the data. The difference ${\rm \Delta = AIC_2-AIC_1}=\chi^2_2-\chi^2_1+2k_2-2k_1$ determines the extent to which model 1 is favored over model 2. The relative probability that model 1 is statistically preferred is given by \begin{equation}
P = \frac{\rm{exp}(-\rm{AIC}_1/2)}{\rm{exp}(-\rm{AIC}_1/2)+\rm{exp}(-\rm{AIC}_2/2)}=
\frac{1}{1+\rm{exp}(-\Delta/2)}\,.
\end{equation}

\section{Results} \label{sec:results}


The left panel of Figure 1 shows the spectra and the best fit models for RX J1713.7$-$3946,
CTB 37B, and CTB 37A. The corresponding model parameters are indicated by the dashed lines
in the right panel. Table 1 gives the statistic-means and the corresponding errors for these
parameters. Note that the best fit model parameters usually are slightly different from the
statistic-means. Here for better comparing with previous studies, the rigidity parameters
have been converted into the corresponding kinetic energy $E=[(Rqc)^2+m^2c^4]^{1/2}-mc^2$,
where $m$ is the rest mass of the particle.
The normalization of the proton distribution $N_{0\rm, p}$ has been converted into the
total energy content of protons, $W_p$, above 1 GeV for a given density and distance.
The total energy content in electrons with $E>1$ GeV is indicated with $W_e$. $W_B$ is
the energy content in the magnetic field assuming a volume filling factor of unity, which
should be considered as an upper limit since radio observations show that the magnetic
field has filamentary structure and concentrates on a shell near the shock front \citep{Reach2005}.

\begin{figure}
\plottwo{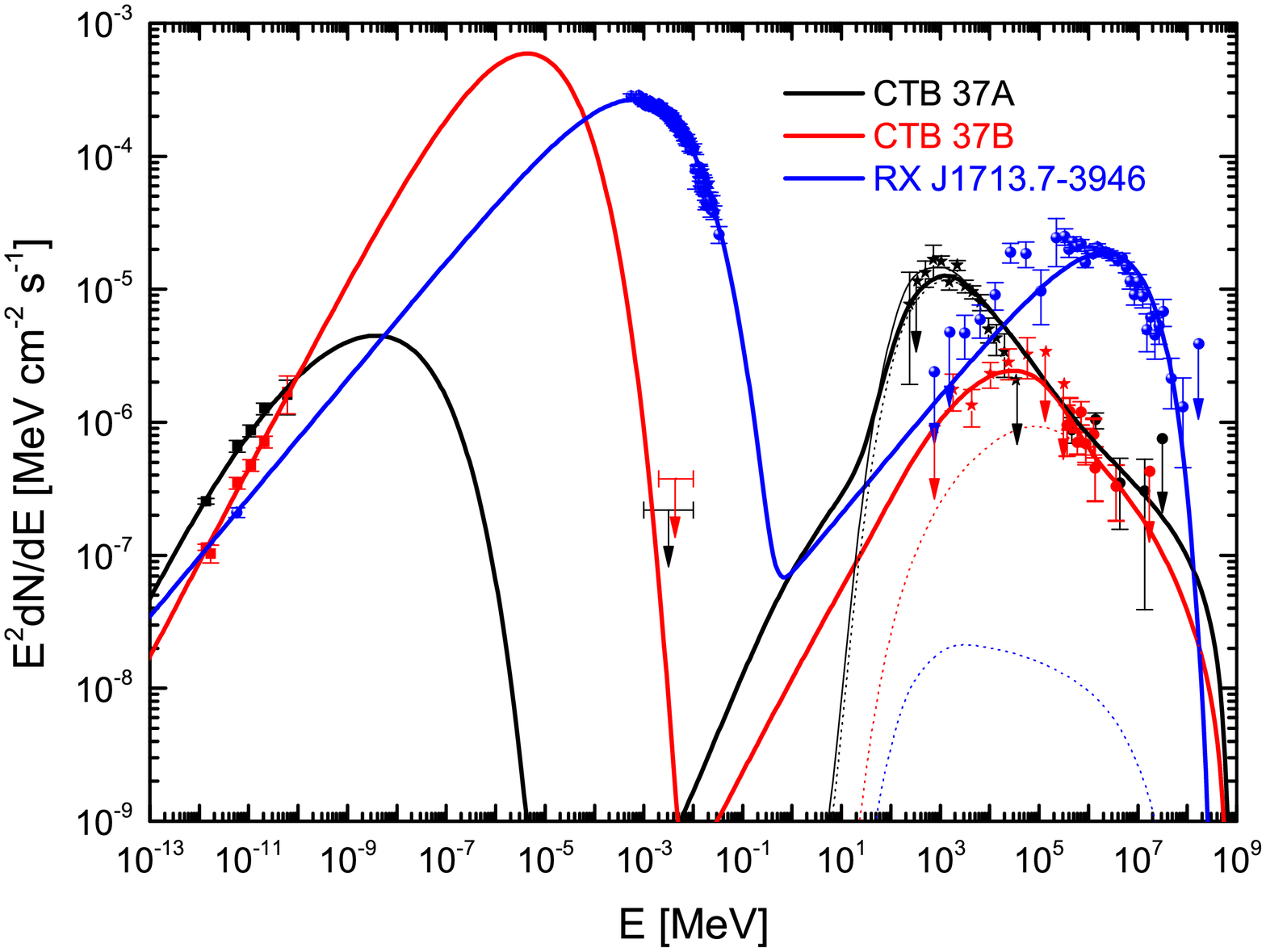}{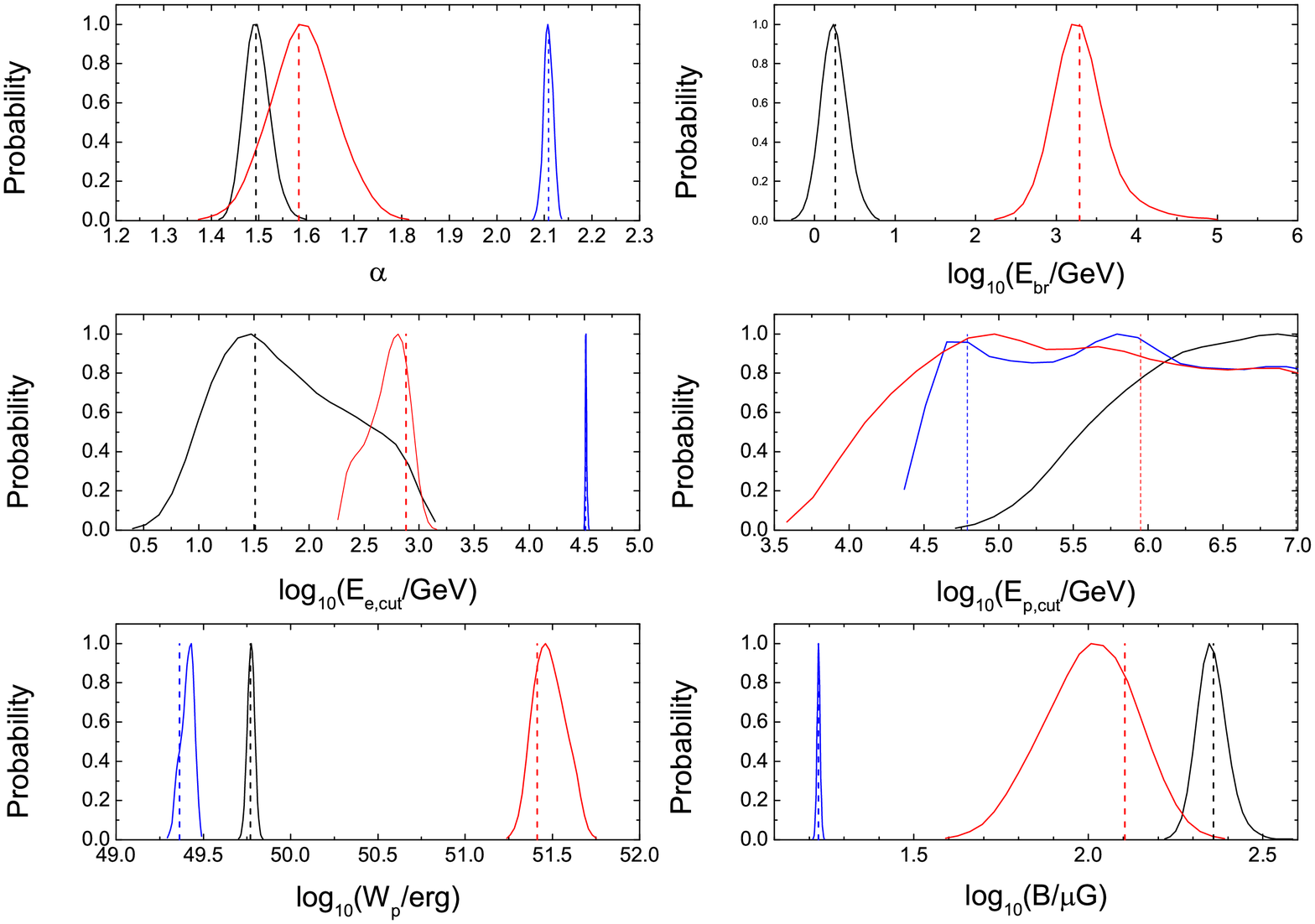}
\caption{
Left: The best fit model spectra for CTB 37A (Black line), CTB 37B (Red line),
and RX J1713.7-3946 (Blue line). The corresponding model parameters are indicated by
dashed lines in the right panel. For CTB 37B and RX J1713.7$-$3946, these spectra correspond
to models with a gas density of n=0.5 cm$^{-3}$ and n=0.01 cm$^{-3}$, respectively.
The dotted lines show the hadronic component for these sources. The thin solid line for
CTB 37A shows the $\gamma$-ray spectrum produced by protons with the high-energy power-law
distribution extended to low energies without a spectral break.
Right: 1-dimensional probability distribution of the model parameters. The vertical dashed
lines indicate values for the best-fit model parameters.}
\label{fig1}
\end{figure}

\subsection{RX J1713.7$-$3946}

SNR RX J1713.7$-$3946 is a well-studied TeV bright shell-type SNR. Its emission spectrum is
dominated by radiation of energetic particles. Only recently, very weak thermal X-ray emission
has been detected from the inner region of the remnant \citep{Katsuda15}.
The spectra observed in radio, X-ray, GeV, and TeV bands, which are shown in Figure 1,
are obtained from \cite{Acero2009,Tanaka2008,Federici2015},
and \cite{Aharonian2011}, respectively. The nature of its $\gamma$-ray emission is still
being debated \citep{Fukui2012,Federici2015}. The dominance of the X-ray emission by
energetic electrons via the synchrotron process and the hard GeV spectrum favors a leptonic
model for the $\gamma$-emission (e.g., \citealp{Yuan2012}). The detection of molecular clouds
and high density gas within/surrounding this remnant suggests that hadronic processes
may also contribute to the $\gamma$-ray emission significantly \citep{Sano2010,Sano2015,Inoue2012}.
Recent detailed studies (e.g. \citealp{Sano2010,Sano2013,Sano2015,Fukui2012}) show that
RX J1713.7$-$3946 resides in an environment rich in molecular and atomic gases and there is a
good spatial correspondence between TeV $\gamma$-ray and the interstellar gas, suggesting a
hadronic origin for the $\gamma$-ray emission. However, assuming ionization equilibrium for
the background plasma, \cite{Yuan2011} systematically investigated the
parameter space of uniform one-zone emission models for the multi-wavelength emission of
SNR RX J1713.7$-$3946, and concluded that the mean gas density should be less than 0.03 cm$^{-3}$,
which is similar to the value of $< 0.02$ cm$^{-3}$ derived from X-ray spectral analyses \citep{Cassam2004}.
Given uncertainties in the shock cloud interaction and cosmic ray interaction with dense molecular clouds,
we consider two effective values for the mean density of the emission region: 0.01 cm$^{-3}$ and
1.0 cm$^{-3}$. An even higher value for the effective density will lead to prominent GeV emission
via the hadronic process, therefore poor fit to the overall spectrum, and reduce the energy content
in relativistic protons.
Here, we also adopt a distance of $1$ kpc, and a radius of $\sim10$ pc for evaluation of the energy contents.

Given the high quality of the X-ray and $\gamma$-ray data and the relatively simple model proposed here, the model parameters are well-constrained and consistent with previous studies in the context of leptonic emission for the $\gamma$-rays
\citep{Fan2010,Yuan2011, Acero2015}.  
For n=0.01 cm$^{-3}$, the magnetic field obtained by us ($17.0 \pm 0.2~\mu$G) is slightly higher
than that derived by Yuan et al. (2011) ($12 ~\mu$G). Such a differences may be attributed to the
high energy density of 1 eV cm$^{-1}$ adopted for the IR background photons and a 19\% scaled
down of the TeV fluxes \citep{Aharonian2011}. In any case, the obtained magnetic field is much
lower than local values (of the order of mG) inferred from variability of X-ray filaments \citep{Uchiyama2007,Inoue2012}
and the value of $\sim 100\mu$G in the hadronic scenario for the $\gamma$-ray emission \citep{Sano2015,Zirakashvili10}.
The synchrotron cooling time of 1340 years at $E_{e\rm, cut}=32$ TeV, which is about 2 times lower
than the upper limit given by equation (4) for a shock speed of 3000 km/s, is slightly lower the age
of 1600 years commonly adopted in literature \citep{Acero2009}. A broken power-law model for electrons
with synchrotron cooling time at the break energy equal to the age will not change other model parameters significantly.
The high-energy cutoff of the proton distribution is not well constrained.
A distinct spectral component above 100 TeV is expected if the high-energy cutoff exceeds $10^{15}$ eV,
which can be tested with future observations.
The total energy content of relativistic protons is $2.6 \times 10^{49}$ erg, which is reasonable for
a relatively young remnant, and the energy contents in electrons above 1 GeV and in the magnetic field
are comparable as shown in earlier studies \citep{Liu2008, Yang2013}. For n = 1.0 cm$^{-3}$
(green lines in Figure 2 and the second row in Table 1 for RX J1713.7$-$3946), the model parameters
have marginal changes except that the high-energy cutoff of the proton distribution can also be
constrained to be 83 TeV.

Contributions to $\chi^2$ from data in the radio, X-ray, GeV, and TeV
bands for the two models are, respectively, 0.008, 380, 32.8, and 96.1 for $n=0.01$ cm$^{-3}$ and 0.07, 381, 20, and 81.2 for $n=1.0$ cm$^{-3}$. The X-ray data dominate the value of $\chi^2$ indicating that the one zone model with an exponential cutoff for the high-energy electron distribution may not give a sufficient description to the spatially integrated X-ray spectrum \citep{Tanaka2008, Fan2010}.
Although compared with the model with $n =0.01$ cm$^{-3}$, the improvement of the spectral model with $n = 1.0$ cm$^{-3}$ is significant with a relative probability for the former of $\sim10^{-6}$, the $\chi^2$ in the GeV and TeV bands for $n = 1.0$ cm$^{-3}$ are still quite high. Recently \cite{Abdalla2016} carried out detailed $\gamma$-ray spectral analysis of RX J1713.7$-$3946 and found that the combined GeV and TeV spectra favor a broken power-law distribution in both the leptonic and hadronic scenarios.  Although the spectral fit may be improved significantly if we adopt a broken power-law distribution, the synchrotron cooling time at the break energy of a few TeV is at least 10 times longer than the age of the remnant (1600 yr). As pointed out by \cite{Abdalla2016}, the temporal evolution of SNR may play a role in the formation of such a broken power-law spectrum, which should also improve the X-ray spectral fit \citep{Ellison2012, Fan2013}. Further exploration of this scenario will be presented in a separate paper.


\begin{figure}
\plottwo{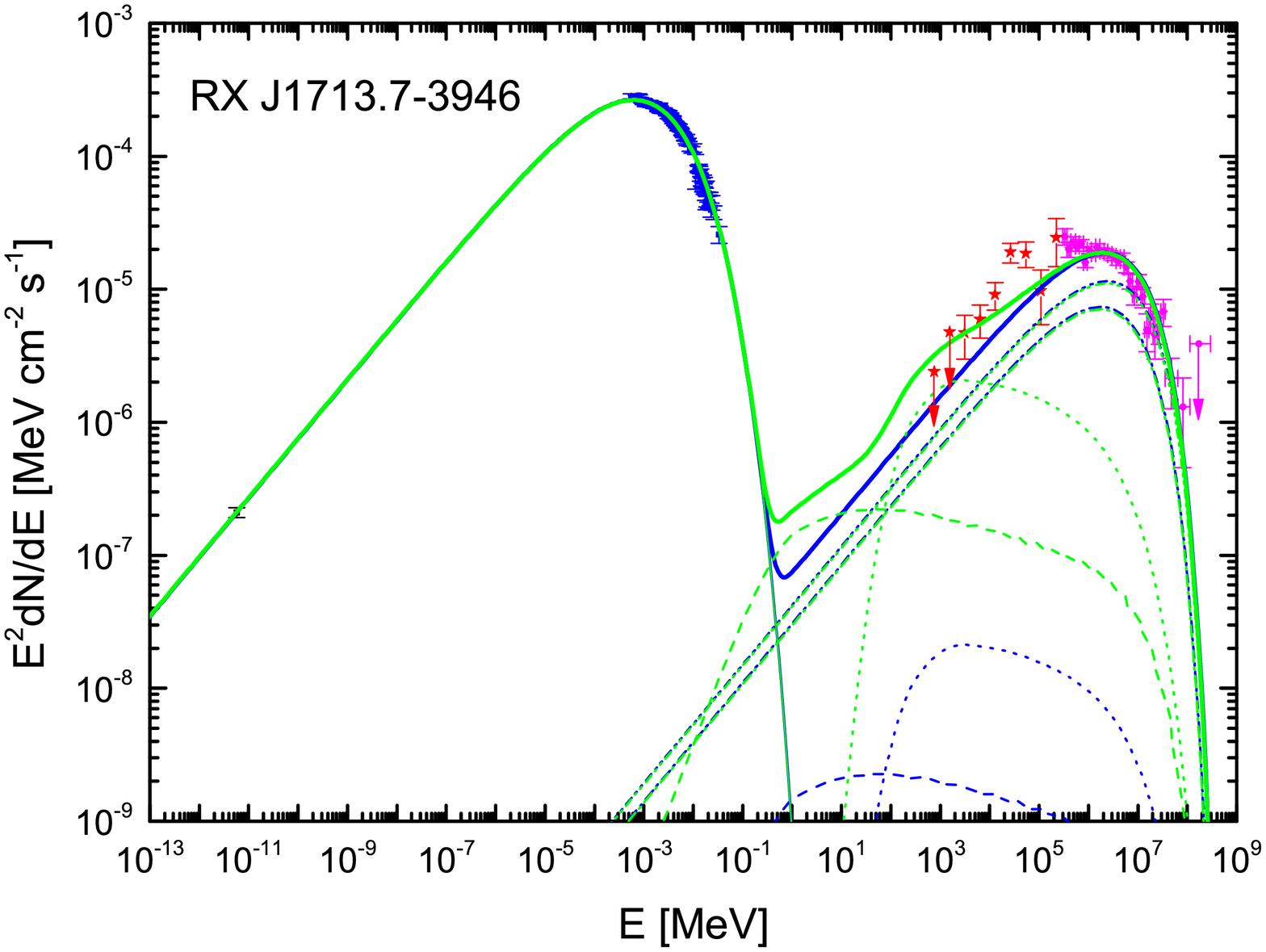}{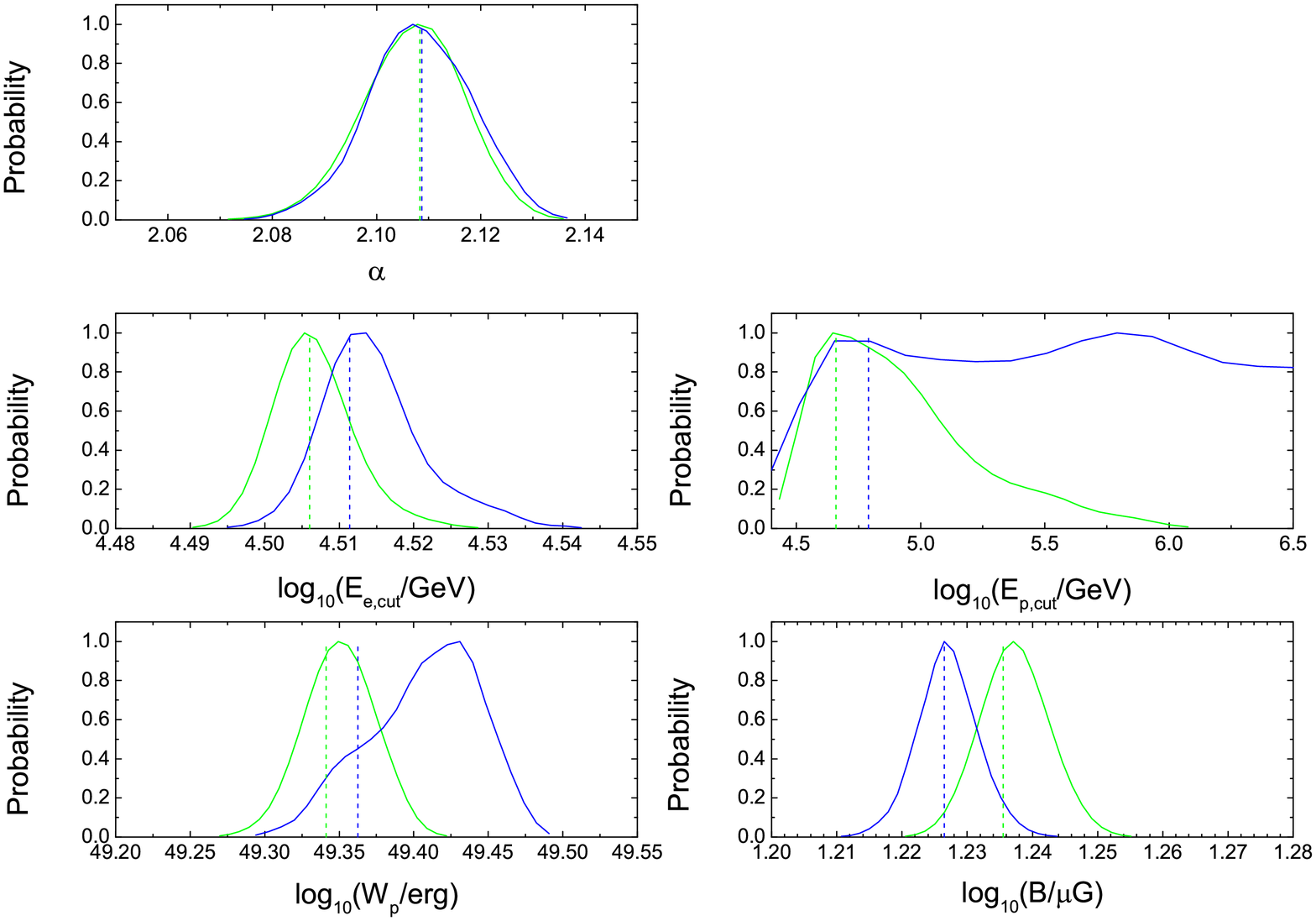}
\caption{
Left: Best fit spectra for RX J1713.7$-$3946 with two different values for the density of
the background plasma. The blue lines are for the same model as shown in Figure 1. The green
lines correspond to a model with a higher density (n= 1 cm$^{-3}$). The dashed lines are
for the bremsstrahlung emission. The dashed-dotted lines and the dashed-dotted-dotted lines are for IC of the CMB and infrared photon fields, respectively. Right:
1-dimensional probability distribution of the model parameters. The vertical dashed lines
indicate the best-fit values.
\label{fig2}}
\end{figure}

\begin{table}[h]
\addtolength{\tabcolsep}{-3pt}
\caption{Spectral fitting parameters. \label{tbl-1}} 
\tiny
\centering 
\begin{tabular}{l|c|c|c|c|c|c|c|c}
\hline\hline 
\raisebox{1.5ex}{Source Name} & \raisebox{1.5ex}{$\alpha$} &\raisebox{1.5ex}{log$_{10} {E_{\rm br}\over{\rm GeV}}$}
& \raisebox{1.5ex}{log$_{10} {E_{\rm e, cut}\over{\rm GeV}}$} & \raisebox{1.5ex}{log$_{10} {E_{\rm p, cut}\over{\rm GeV}}$}
& \raisebox{1.5ex}{${B\over\mu{\rm G}}$} & \raisebox{1.5ex}{log$_{10} {W_{p}\over{\rm erg}}$}
& \raisebox{1.5ex}{${W_{B}\over W_{e}}$}&  Reduced $\chi^2$
\\ 
&&&&&&&& for the best-fit \\
\hline
 & & & & &  & $49.41_{-0.04}^{+0.04}$+ & $5.46 \times$ & $\frac{509}{240-5}=$  \\
\raisebox{1.5ex}{RX J1713.7$-$3946}
& \raisebox{1.5ex}{$2.11_{-0.01}^{+0.01}$}
& \raisebox{1.5ex}{NA}
& \raisebox{1.5ex}{$4.51_{-0.006}^{+0.005}$}
& \raisebox{1.5ex}{$>4.51$}
& \raisebox{1.5ex}{$1.23_{-0.004}^{+0.004}$}
& log$_{10}\left[\left({\rm n}\over{0.01{\rm cm}^{-3}}\right)^{-1}\left({{\rm D}\over 1{\rm kpc}}\right)^2\right]$
& \raisebox{1.5ex}{$\left({{\rm D}\over 1{\rm kpc}}\right)^{-2}\left({{\rm R}\over 10{\rm pc}}\right)^3$}
& \raisebox{1.5ex}{$2.17$}
 \\ 
 \hline
 & & & &  &  & $49.35_{-0.02}^{+0.02}$+ & $5.72 \times$ & $\frac{482}{240-5}=$  \\
\raisebox{1.5ex}{RX J1713.7$-$3946}
& \raisebox{1.5ex}{$2.11_{-0.01}^{+0.01}$}
& \raisebox{1.5ex}{NA}
& \raisebox{1.5ex}{$4.51_{-0.005}^{+0.005}$}
& \raisebox{1.5ex}{$4.92_{-0.31}^{+0.31}$}
& \raisebox{1.5ex}{$1.24_{-0.005}^{+0.005}$}
& log$_{10}\left[\left({\rm n}\over{1.0{\rm cm}^{-3}}\right)^{-1}\left({{\rm D}\over 1{\rm kpc}}\right)^2\right]$
& \raisebox{1.5ex}{$\left({{\rm D}\over 1{\rm kpc}}\right)^{-2}\left({{\rm R}\over 10{\rm pc}}\right)^3$}
& \raisebox{1.5ex}{$2.05$}
 \\ 
 \hline
 & & & & &  & $51.49_{-0.09}^{+0.10}$+ & $32.9 \times$ & $\frac{13.7}{19-6}=$  \\
\raisebox{1.5ex}{CTB 37B}
& \raisebox{1.5ex}{$1.59_{-0.07}^{+0.07}$}
& \raisebox{1.5ex}{$3.31_{-0.31}^{+0.29}$}
& \raisebox{1.5ex}{$2.70_{-0.21}^{+0.18}$}
& \raisebox{1.5ex}{$>4.57$}
& \raisebox{1.5ex}{$2.00_{-0.13}^{+0.13}$}
& log$_{10}\left[\left({\rm n}\over{0.5{\rm cm}^{-3}}\right)^{-1}\left({{\rm D}\over 13.2{\rm kpc}}\right)^2\right]$
& \raisebox{1.5ex}{$\left({{\rm D}\over 13.2{\rm kpc}}\right)^{-2}\left({{\rm R}\over 20{\rm pc}}\right)^3$}
& \raisebox{1.5ex}{$1.05$}
 \\ 
  \hline
 & & & & & & $50.47_{-0.04}^{+0.04}$+ & $1.75 \times 10^{4} \times $ & $\frac{14.3}{19-6}=$  \\
\raisebox{1.5ex}{CTB 37B}
& \raisebox{1.5ex}{$1.51_{-0.11}^{+0.11}$}
& \raisebox{1.5ex}{$2.45_{-0.34}^{+0.36}$}
& \raisebox{1.5ex}{$1.57_{-0.78}^{+0.79}$}
& \raisebox{1.5ex}{$>4.60$}
& \raisebox{1.5ex}{$2.71_{-0.15}^{+0.14}$}
& log$_{10}\left[\left({\rm n}\over{10.0{\rm cm}^{-3}}\right)^{-1}\left({{\rm D}\over 13.2{\rm kpc}}\right)^2\right]$
& \raisebox{1.5ex}{$\left({{\rm D}\over 13.2{\rm kpc}}\right)^{-2}\left({{\rm R}\over 20{\rm pc}}\right)^3$}
& \raisebox{1.5ex}{$1.10$}
 \\ 
  \hline
 & & & & &  & $49.77_{-0.02}^{+0.02}$+ & $575 \times$ & $\frac{33.8}{21-6}=$  \\
\raisebox{1.5ex}{CTB 37A}
& \raisebox{1.5ex}{$1.50_{-0.03}^{+0.03}$}
& \raisebox{1.5ex}{$0.25_{-0.16}^{+0.16}$}
& \raisebox{1.5ex}{$1.80_{-0.61}^{+0.67}$}
& \raisebox{1.5ex}{$>6.18$}
& \raisebox{1.5ex}{$2.36_{-0.04}^{+0.04}$}
& log$_{10}\left[\left({\rm n}\over{100{\rm cm}^{-3}}\right)^{-1}\left({{\rm D}\over 7.9{\rm kpc}}\right)^2\right]$
& \raisebox{1.5ex}{$\left({{\rm D}\over 7.9{\rm kpc}}\right)^{-2}\left({{\rm R}\over 10{\rm pc}}\right)^3$}
& \raisebox{1.5ex}{$2.25$}
 \\ 

\hline 
\end{tabular}
\tablecomments{Errors are for 1$\sigma$ statistical uncertainties; Lower limits
correspond to 1$\sigma$ confidence level; ``NA'' represents that the distribution is
approximately a single power-law with an exponential cutoff.}
\end{table}

\subsection{CTB 37A}

The radio data for CTB 37B and CTB 37A  are taken from \citet{Kassim1991}.
SNR CTB 37A has a partial shell with an extended breakout to the
south \citep{Kassim1991}.
The distance to CTB 37A has been estimated from 21-cm absorption
measurements to be 10.3 $\pm$ 3.5 kpc by \cite{Caswell1975}.
Recently, \cite{Maxted2013} and \cite{Tian2012} pointed out that CTB 37A and
the $\sim 60 \rm km s^{-1}$ 1720 MHz OH masers in the direction should locate
in the inner 3 kpc of the Galaxy, which implies a CTB 37A distance between 6.3
and 9.5 kpc. We will adopt a distance of 7.9 kpc in this paper.

CTB 37A does not have diffuse nonthermal X-ray emission.
Using Chandra and XMM-Newton observations, \citet{Aharonian2008a} found diffusive
thermal X-ray emission from the east part of the remnant and derived a flux
upper limit of $3.5 \times 10^{-13}$ erg cm$^{-2}$s$^{-1}$ for nonthermal emission
in the 1-10 keV range. \cite{Sezer2011} suggested that the diffuse thermal X-ray
emission of CTB 37A is produced by shocked interstellar/circumstellar medium and estimated
an ambient gas density of $\sim 1~f^{-1/2}$ cm$^{-3}$ and an age of
$\sim 3 \times 10^4 ~ f^{1/2}$ yr, where $f$ is the volume filling factor of the
emitting plasma. Assuming a distance of 11.3 kpc, \cite{Pannuti2014} obtained similar
results with an electron density $n_e=0.77$ cm$^{-3}$ and an age of $(3.2-4.2) \times 10^{4}$ yr
by analyzing the X-ray spectrum.
However the remnant is clearly interacting with molecular clouds.
From velocity measurements of molecular clouds associated with the remnant, \cite{Reynoso2000}
estimated the masses of individual clouds ranging from $1.3 \times10^3$ $M_\odot$ and
$5.8 \times 10^{4}$  $M_{\odot}$ with $H_2$ densities between 150 $\rm cm^{-3}$
and 660 $\rm cm^{-3}$ assuming a distance of 11.3 kpc, which implies a size of the remnant close to 28 pc.
Based on the CS(1-0) emission, \cite{Maxted2013} obtained a density of $(3-10) \times 10^3$
cm$^{-3}$ and a mass of $(500-2300) M_\sun$, which is consistent with the estimation given by \cite{Reynoso2000}.
We will assume an effective density of 100 cm$^{-3}$ for calculation of the $\gamma$-ray emission
and a radius of 10 pc for estimation of the energy content in the magnetic field.

HESS observations show that the TeV fluxes from CTB 37A and CTB 37B are comparable and
CTB 37A has a slightly harder TeV spectrum than CTB 37B \citep{Aharonian2008a}. The GeV flux of
CTB 37A is much higher than that of CTB 37B \citep{Brandt2013}. Using the Pass 8 data from the \emph{Fermi}-LAT,
we reanalysized the GeV spectrum of CTB 37A \citep{Xin2016} and found that the GeV fluxes are
systematically lower than those given by \citet{Brandt2013} due to identification of more sources
nearby, and the overall $\gamma$-ray spectrum is consistent with a single power law model. However,
compared with the radio spectrum of CTB 37B, the radio spectrum of CTB 37A shows clear evidence of
softening toward high frequencies (Figure 1). We will adopt a broken power law model to fit the overall spectrum.

The inferred break energy is about 2 GeV (Table 1). Since the $\gamma$-ray emission is dominated by
decay of $\pi^0$ for the high density adopted here, a broken power-law distribution with such a
low break energy produces a $\gamma$-ray spectrum (black dotted line in Figure 1) almost identical
to the $\gamma$-ray spectrum produced by protons with a single power-law distribution with a spectral
index of 2.5 (thin black line in Figure 1). Therefore a broken power law distribution of high-energy
protons is not necessary to reproduce the observed $\gamma$-ray spectrum. However,
the radio spectrum justifying the broken power-law distribution. The synchrotron cooling time at the break
(2 GeV) and the cutoff (63 GeV) energy of electrons are about $10^5$ and 4000 years, respectively, which
are compatible with the age ($\sim 3\times 10^4$ years) of the hot plasma in the east part of the remnant. The fact that the synchrotron
cooling time at the cutoff energy is more than 2 times shorter than the age of the remnant ($\sim 3 \times 10^4$ years) demands
continuous electron acceleration in the late stage of the SNR evolution when the shock is interacting with molecular clouds.

We also infer a very hard distribution below 2 GeV, which is harder than even a strong shock produces at the shock front.
This may be attributed to either compression of Galactic cosmic rays or Coulomb loss processes as suggested
by \citet{Chevalier1999} or to spatial transport processes away from the shock.
The latter possibility will be discussed in a future publication (in preparation).
We note that simple compression cannot change the power-law index. Also, the total
energy content of protons with $E>1$ GeV is about $10^{50}$ erg, which is consistent
with the scenario of SNR origin for the Galactic cosmic rays, but more than 2 orders
of magnitude higher than the energy content of Galactic cosmic rays within the shell
of the remnant assuming an energy density of $\sim 1$ eV cm$^{-3}$. Simple compression
of Galactic cosmic rays cannot reproduce the observed nonthermal emission.

A strong mean magnetic field of $\sim 230~\mu$G  is inferred, which can be
attributed to shock interaction with molecular clouds. Indeed, strong magnetic fields
($0.2-1.5$ mG) in post-shock gas are estimated by \cite{Brogan2000} based on observations
of OH (1720 MHz) masers from shocked molecular clouds. These fields are much higher than
the value of 20 $\mu$G field inferred by \cite{Brandt2013} assuming a leptonic origin for the GeV
emission. Such a strong magnetic field also leads to a higher energy content in the
magnetic field than that in energetic electrons, as has been noticed in remnants slightly
older than RX J1713.7$-$3946  \citep{Yang2014}. The energy contents in the magnetic field and
relativistic protons are actually comparable, which is consistent with the theoretical
expectation for shock interaction with molecular clouds \citep{Blandford1982}.

\cite{Castro2010} reported the detection of $\gamma$-ray emission of CTB 37A using data
from \emph{Fermi}-LAT, and found that the GeV spectrum can be fitted by a simple power-law model with a spectral
index $\Gamma=-2.19 \pm 0.07$. They also found that a harder spectrum with an index of $-1.46 \pm 0.32$ and a high-energy cutoff of 4.2 GeV gives a slightly improved fit to the \textit{Fermi} data.
\cite{Brandt2013} modeled the multi-wavelength spectrum of CTB 37A using a combination of leptonic
and hadronic emission. The leptonic component has a power-law distribution with a spectral index of $-1.75$ and
a high-energy cutoff of 50 GeV; the hadronic component has a power-law distribution with an index of $-2.3$.
The GeV emission is dominated by the bremsstrahlung process while the TeV flux is dominated by hadronic emission.
With the newly released \textit{Fermi} data, we found that both the GeV and TeV emission can be attributed to energetic protons, and the model we have for the multi-wavelength spectrum is quite different
from that proposed by \cite{Brandt2013}.
However, we notice that the value of reduced $\chi^2$ of the best-fit model (2.25) is high with contributions to the $\chi^2$ from the \textit{Fermi} (22.7) and HESS (8.7) data dominant. Our best fit model has a $\gamma$-ray spectral index of $-2.5$, which is lower (softer) than that for the TeV spectrum and higher (harder) than that for the GeV spectrum. If future observations confirm $\gamma$-ray spectral hardening toward higher energies, which may be attributed to nonlinear diffusive shock acceleration \citep{Ellison2012} or interaction of cosmic ray with molecular clouds \citep{Gabici2014}, the simple spectral model proposed here will need to be revised to take into account the relevant physical processes.

\subsection{CTB 37B}

Since no prominent diffuse nonthermal X-ray emission is detected from CTB 37B, X-ray
flux above 2 keV is treated as an upper limit for the nonthermal emission following \citet{Xin2016},
who also gave the GeV data using the recently released Pass 8 data of the \emph{Fermi}-LAT.
It is interesting to note that the GeV spectrum appears to have a peak near 10 GeV.
The spectrum of the diffuse thermal X-ray emission implies a low density of $\sim 1$ cm$^{-3}$
and an age of $\sim$ 5000 years \citep{Aharonian2008b}, which are consistent with a pre-shock
electron density of $0.4 \pm 0.1$ cm$^{-3}$ estimated by \cite{Nakamura2009} with a non-equilibrium
collisional ionization model for Suzaku and Chandra observations.
The distance to CTB 37B has been estimated to be $10.2 \pm 3.5$ kpc by \cite{Caswell1975},
$\approx 8$ kpc by \citet{Green2009} and a large value of $\sim$ 13.2 kpc by \citet{Tian2012}.
We adopt the most recent distance estimate of 13.2 kpc in this paper.

Due to the absence of the synchrotron X-ray emission, \citet{Aharonian2008b} argued that the TeV
emission should have a hadronic origin. A hadronic model will require a spectral break near a
few hundreds of GeV to produce a $\gamma$-ray spectral peak near 10 GeV. Synchrotron X-ray emission
is expected if the electron distribution extends into the TeV range. Therefore the electron
distribution should be a single power law with a high-energy cutoff below 1 TeV while the proton
distribution follows a broken power law with a spectral break at a few hundreds of GeV. With a
background gas density of $0.5$ cm$^{-3}$, \citet{Xin2016} found that the energy content of relativistic
protons exceeds 10$^{51}$ erg in the hadronic scenario. Even in the leptonic scenario for
the $\gamma$-ray emission, they found that the energy content in electrons above 1 GeV is about
one order of magnitude higher than that of other $\gamma$-ray remnants. Due to introduction of
high-energy cutoffs below 10 TeV, their models produce TeV spectra much softer than the
observed one. Adopting the same density for the background plasma, our best fit model results
show a hybrid origin for the $\gamma$-ray emission with the hadronic and the IC process having
comparable contribution in the GeV range (red lines in Figure 3). The TeV emission however is
dominated by the hadronic process as suggested by \citet{Aharonian2008b}. Due to contribution
to the GeV emission via the IC process, the break energy of the proton distribution is about 20
TeV leading to a spectral peak near 100 GeV for the hadronic emission component (red dotted line
in Figure 1). The model also produces a TeV spectrum slightly softer than that for CTB 37A,
which is consistent with HESS observations \citep{Aharonian2008a}.

We infer a strong magnetic field of $100\ \mu$G and a very high energy content of $\sim
3\times 10^{51}$ erg in relativistic protons, which are consistent with the results of
\citet{Xin2016}. The energy content in the magnetic field is slightly higher than that
in GeV electrons compatible with the evolution of younger remnants \citep{Yang2014, Guo2012}.
The high-energy cutoff of the electron distribution is about $500$ GeV which is mostly
constrained by the observed GeV emission via the IC process. The synchrotron cooling time
at the cutoff energy of electrons is about 2800 years, which is about two times shorter
than the age estimate ($\sim 5000$ years). Taking into account the effect of synchrotron loss on the electron distribution, one may adopt a broken power-law distribution for electrons with a break energy below 200 GeV for the magnetic field inferred above. Such a model may reproduce
the observed spectrum if one includes optical photons in the background, which may
contribute to the GeV $\gamma$-ray emission significantly. We will not explore this
model in details since the model parameters will not change dramatically.

We also infer a very hard distribution with an index of 1.6 from hundreds of MeV to
hundreds of GeV for electrons and to tens of TeV for protons.
Such a distribution cannot be attributed to compression of Galactic cosmic rays or
Coulomb loss processes. Nonlinear diffusive shock acceleration will not produce a
single power-law distribution with a very hard spectrum. Stochastic particle
acceleration by compressible turbulence in the downstream of the shock may be
able to account for such a hard spectrum \citep{Ostrowski1999, Bykov2000, Liu2008, Fan2013}.
As mentioned above, spatial transport of accelerated particles away from the shock front will
also produce a harder spectrum.


\begin{figure}
\plottwo{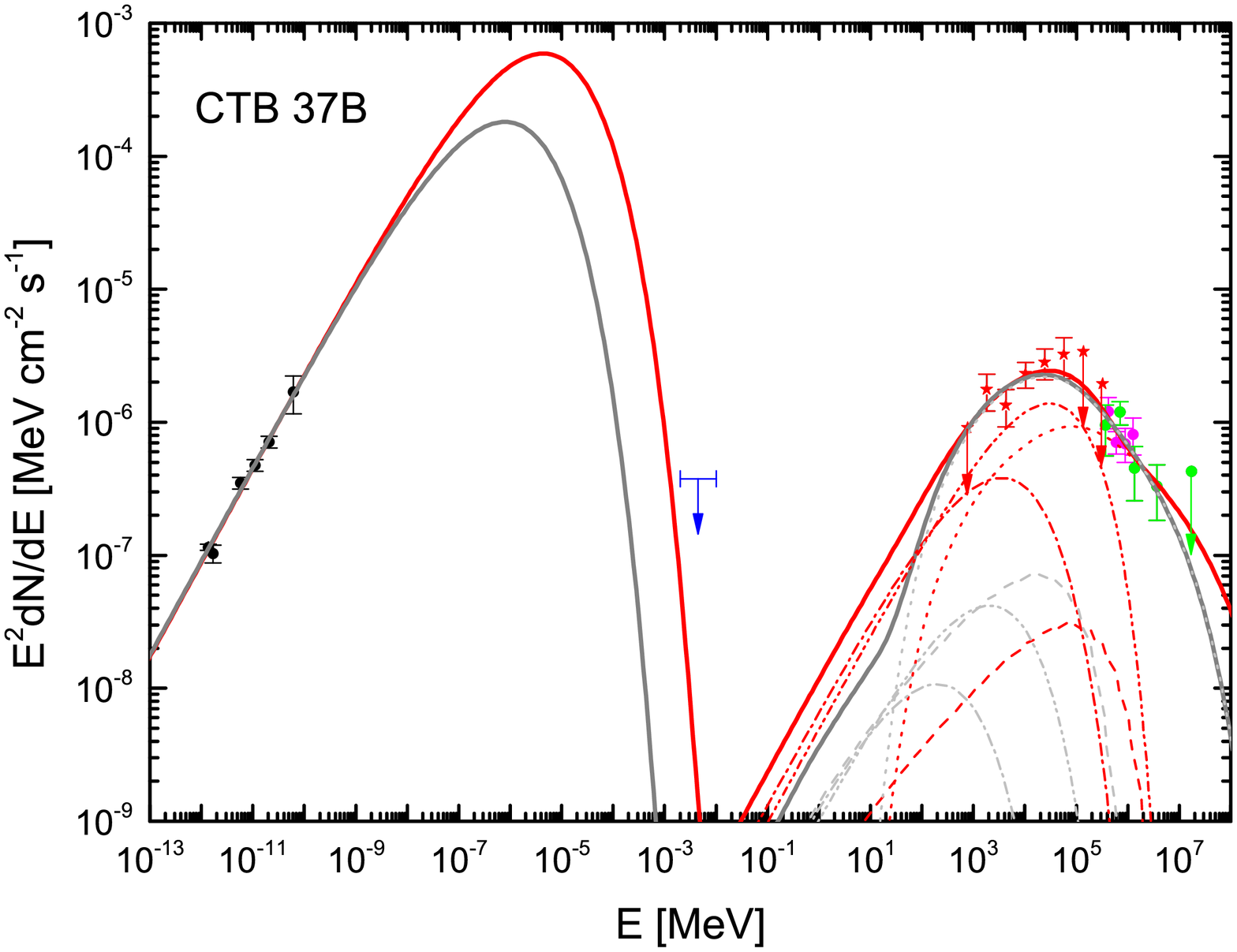}{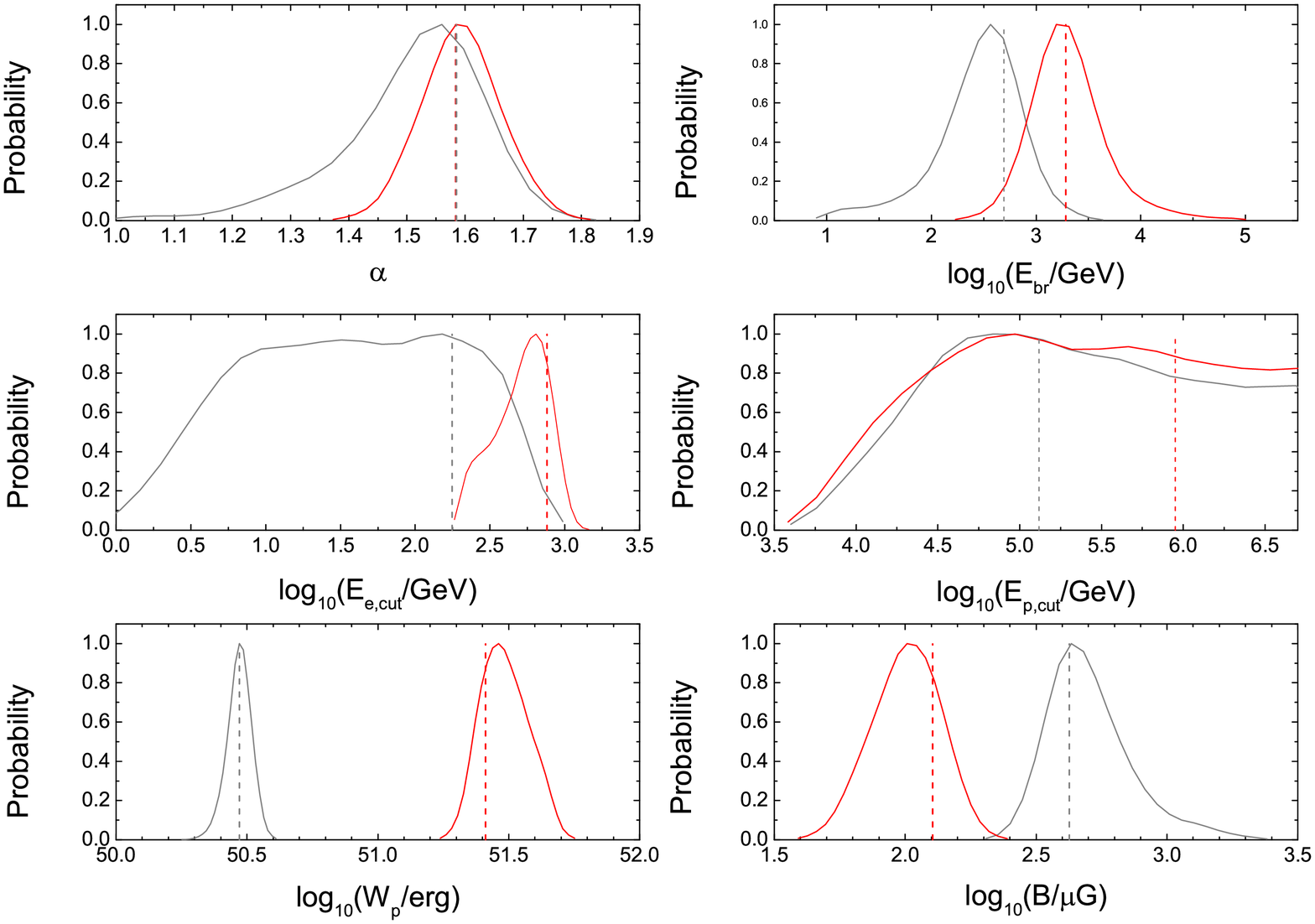}
\caption{
Left: Best fit spectra for CTB 37B with two different values for the density of the
background plasma. The red lines are for the same model shown in Figure 1. The gray
lines correspond to a model with a higher density of n=10 cm$^{-3}$. The dashed
lines are for the bremsstrahlung emission. The dash-dotted lines and the dash-dotted-dotted
lines are for IC of the CMB and infrared photon fields, respectively.
Right: 1-dimensional probability distribution of the model parameters. The vertical dashed
lines indicate the best-fit values.
\label{fig3}}
\end{figure}

We notice that there is increasing evidence that CTB 37B may be interacting with molecular
clouds \citep{Frail1996, Tian2012, Jiang2015}. The broken power-law distribution of relativistic
protons also favors shock interaction with molecular clouds.
By adopting an effective density of $10$ cm$^{-3}$, the $\gamma$-ray emission is then dominated
by the hadronic process and the energy content of relativistic protons is reduced by more than
one order of magnitude, which is comparable to other $\gamma$-ray remnants 
(See gray lines in Figure 3 and the second row in Table 1 for CTB 37B). The break energy is reduced to about 300 GeV
in this case leading to a $\gamma$-ray spectral peak near 10 GeV. However, a stronger magnetic
field of 0.5 mG is needed to reproduce the radio fluxes for the decrease of $N_{0, e}$, which
also suppresses contributions to $\gamma$-rays by the IC process. For a volume filling factor
of a few percent, which is compatible with radio observations, the energy content of the magnetic
field will be comparable to that of relativistic protons.
The synchrotron cooling time of electrons at the cutoff energy of $40$ GeV is about 1900 years,
which is about 2 times shorter than the age estimate ($\sim 5000$ years), implying continuous particle acceleration.

A harder distribution with an index of 1.51 is inferred due to the decrease of the cutoff energy
for the electron distribution. Such a distribution also challenges the diffusive shock particle
acceleration theory and may imply continuous particle acceleration downstream of the shock
\citep{Bykov2000, Liu2008, Malkov11}. Spatial transport from the shock may also harden the
spectrum (Jokipii and Liu, in preparation). In W44, \citet{Uchiyama2010} shows that the hard
radio spectrum may be attributed to secondary electrons and positrons produced in the pp elastic
collisions. The radio emission of CTB 37B cannot be attributed to secondary electrons and positrons
since the radio luminosity is comparable to the $\gamma$-ray luminosity.
Compared with the model (red lines in Figure 3) with a lower effective density of 0.5 cm$^{-3}$, this model (grey lines in Figure 3) with an effective density of 10 cm$^{-3}$
reduces the overall energy content to a value of $\sim 10^{50}$ erg compatible with typical SNRs and in agreement with the SNR origin of Galactic cosmic ray. A reduced $\chi^2$ of $\sim 1$ shows that both models give equally good fit to the broadband spectra. An even higher value for the effective density will lead to a
stronger magnetic field and less energy content in relativistic protons, which requires a smaller
volume filling factor for the magnetic field and relativistic protons to maintain energy
equipartition \citep{Blandford1982}.
More observations are needed to confirm shock interaction with molecular clouds in this remnant \citep{Jiang2015}.

Recently, \cite{Xin2016} reported the gamma-ray emission of CTB 37B by using
7-year \emph{Fermi}-LAT data, and found that the multi-wavelength spectrum can be well
fitted with a leptonic or a hadronic model. The parameters of their leptonic
model are similar to ours except for a slightly higher cutoff energy of the electron
distribution for the absence of hadronic contribution to the $\gamma$-rays, which also
leads to a deficit in the model predicted TeV flux. In their hadronic model, electrons
and protons have softer distributions with different spectral indexes. The proton
distribution cuts off at 3 TeV with an energy content much higher than ours for
the adoption of a lower density of the background plasma, which leads to a value of $K_{ep}$
more than one order of magnitude lower than ours. Our magnetic field of 0.5 mG is
more than 2 times higher than their value for the lower energy content of electrons in our model.
In order to estimate the diffuse neutrino flux from SNRs, \cite{Mandelartz2015} use a
similar one-zone model to perform multi-wavelength spectral fit to 22 SNRs. CTB 37A and
CTB 37B are among those sources. However, they adopted a single power-law distribution and
the normalization, spectral index, and cutoff energy of the electron distribution are
independent of those for protons. Their model parameters in general are not as well
constrained as ours and, with the recently released \emph{Fermi} data, we have an up-to-date
GeV flux measurement for both sources.

\section{Conclusion and Discussion} \label{sec:Conclusion and Discussion}

Considering the SNR origin of Galactic cosmic rays and physical processes related
to acceleration and radiation of high-energy particles, we propose a simple one-zone emission model
with at most 6 parameters for nonthermal emission spectrum of shell type SNRs driven by shocks of
supernova explosions. The model assumes that the ratio of the normalization of electron
distribution to that of protons is 0.01 and adopts reasonable values for the effective density of
background plasma in the emission region. With the MCMC algorithm, multi-wavelength spectral
measurement of SNRs can be used to constrain other parameters related to radiation of relativistic
particles in these remnants. Using three shell type SNRs located within 2 degrees on the sky and
having distinct $\gamma$-ray spectra as examples, we carried out detailed spectral fitting and
studied the physical implications of the model parameters.

By fitting SNRs with distinct $\gamma$-ray spectra with a unified model, it is possible to
explore evolution of energetic particle distributions in SNRs. We find that in general the particle
distribution becomes harder with aging of the shock and it is challenging to explain the
hard electron distribution in CTB 37B via the diffusive shock particle acceleration, compression of
Galactic cosmic rays, secondaries of the pp process, and/or via Coulomb loss processes. Continuous
particle acceleration in the shock downstream by compressible turbulence may produce a very hard
distribution \citep{Liu2008}. We also suggest that spatial transport of cosmic rays away from
the shock may account for the hard observed spectrum. The synchrotron energy loss time at the high-energy
cutoff of the electron distribution is always shorter than and appears to increase with the age of
the corresponding remnant, which also requires continuous acceleration and suggests that a time-dependent
approach may be necessary to study the bulk of the particle acceleration in SNRs as is the case
for a stochastic particle acceleration model proposed by \citet{Fan2013}.
These results suggest that, for mature SNRs, the classical diffusive shock acceleration process may not
dominate the overall particle acceleration. Detailed studies of spatial structure of individual
SNRs may be used to separate features associated with shock acceleration from other nonthermal features
to determine the dominant particle acceleration processes \citep{Sano2015, Abdalla2016}.

Our model predicts a correlation between the radio and $\gamma$-ray spectral hardness, which naturally
produces a TeV spectrum in CTB 37A harder than that in CTB 37B. However given the complexity of processes
involved in the formation of electron distribution in the GeV range and the three $\gamma$-ray emission
processes, such a correlation may not be evident in the observed data and detailed spectral fit is needed.
By adjusting the effective density of ions in the background plasma for relativistic proton interaction with,
the overall energy content in relativistic protons can be confined on the order of $10^{50}$ erg, which
agrees with the SNR origin of Galactic cosmic rays. Other model parameters derived with such a constraint
are also compatible with measured characteristics of these SNRs.

For the sake of simplicity, we do not consider the temporal and spatial evolution of SNR, the radiative cooling effect on the distribution of high-energy electrons, the effects of Galactic cosmic ray compression by shocks and Coulomb collisional energy loss on the distribution of relatively low-energy electrons. We also assume that the spectral index increases by one from low to high energies for the broken power-law distribution. For the three SNRs studied here, these simplifications do not seem to affect the main conclusions above. However the relatively high values of the reduced $\chi^2$ for CTB 37A and RX J1713.7$-$3946 suggest that revision of this simple model is necessary to improve the spectral fit. \textit{Fermi} $\gamma$-ray observations
have revealed a variety of spectral shape \citep{Ackermann2013, Acero2016}. Over the past few years, similar models have been applied to study of individual SNRs. The model parameters for W51 derived by \cite{Aleksic2012} are very similar to those for CTB 37A. The $\gamma$-ray spectrum  of young SNR G349.7$+$0.2 \citep{Abramowski2015} has a spectral break above 10 GeV, reminiscence of the $\gamma$-ray spectrum of CTB 37B. Application of a uniform model to these SNRs is essential to probe the evolution of energetic particle distribution in SNRs. It will also guide further development of the model to incorporate more physical processes affecting the evolution of energetic particle distribution in SNRs.



\acknowledgments

We would like to thank the anonymous referee for very helpful comments,
which help to improve the paper significantly. This work is partially
supported by the Strategic Priority Research Program,
the Emergence of Cosmological Structures, of the Chinese Academy of Sciences,
Grant No. XDB09000000, the Key Laboratory of Particle Astrophysics of Yunnan
Province (Grant 2015DG035), and the NSFC grants 11173064, 11233001, and 11233008.
LZ acknowledges partial funding support by the National Natural Science Foundation
of China (NSFC) under grant No. 11433004.

\end{document}